%% file: main.tex
\documentclass[conference, review]{IEEEtran}
\IEEEoverridecommandlockouts
\usepackage[noadjust]{cite}
\usepackage{amsmath,amssymb,amsfonts}
\usepackage{algorithmic}
\usepackage{subcaption}
\usepackage{amsmath}
\usepackage{textcomp}
\usepackage{hyperref}
\usepackage{xcolor}
\usepackage{booktabs,ragged2e, array}
\usepackage{tabularx,booktabs,caption}
\newcolumntype{Z}{>{\raggedright}X}
\usepackage[flushleft]{threeparttable}
\usepackage{graphicx}
\usepackage{url}
\usepackage{cleveref}
\usepackage{balance}
\def\BibTeX{{\rm B\kern-.05em{\sc i\kern-.025em b}\kern-.08em
    T\kern-.1667em\lower.7ex\hbox{E}\kern-.125emX}}

\begin{document}
\title{Analyzing `Near Me' Services: Potential for Exposure Bias in Location-based Retrieval}
%\title{Analyzing `Near Me' Services: Potential Exposure Bias in Local Search and Recommendation}

\newcommand{\ashmi}[1]{\textcolor{orange}{AB:#1}}
\newcommand{\gourab}[1]{\textcolor{green}{GP: #1}}
\newcommand{\linus}[1]{\textcolor{purple}{LD:#1}}
\newcommand{\todo}[1]{\textcolor{blue}{TODO:#1}}

\author{ 
\IEEEauthorblockN{Ashmi Banerjee}
\IEEEauthorblockA{Technical University of Munich}
\and
\IEEEauthorblockN{Gourab K Patro}
\IEEEauthorblockA{IIT Kharagpur}
\and
\IEEEauthorblockN{Linus W. Dietz}
\IEEEauthorblockA{Technical University of Munich}
\and
\IEEEauthorblockN{Abhijnan Chakraborty}
\IEEEauthorblockA{MPI for Software Systems}
}

\maketitle

\input{abstract.tex}
\input{intro.tex}
\input{related.tex}

\input{data_gathering.tex}
\input{EB_in_search_and_reco.tex}
\input{characterizing_EB.tex}

\input{conclusion.tex}

\section*{Acknowledgement}
This research was supported in part by a European Research Council (ERC) Advanced Grant for the project “Foundations for Fair Social Computing”, funded under the European Union’s Horizon 2020 Framework Programme (grant agreement no. 789373). G. K Patro is supported by a Fellowship from Tata Consultancy Services.

% \ashmi{TODO: convert all .pngs to .pdf/.eps}
\balance
\bibliographystyle{IEEEtran}
\bibliography{main}

\end{document}

%% file: abstract.tex
\begin{abstract}
The proliferation of smartphones has led to the increased popularity of location-based search and recommendation systems. Online platforms like Google and Yelp allow location-based search in the form of a \textit{nearby} feature to query for hotels or restaurants in the vicinity. Moreover, hotel booking platforms like Booking[dot]com, Expedia, or Trivago allow travelers searching for accommodations using either their desired location as a search query or near a particular landmark. Since the popularity of different locations in a city varies, certain locations may get more queries than other locations. Thus, the exposure received by different establishments at these locations may be very different from their intrinsic quality as captured in their ratings. 

Today, many small businesses (shops, hotels, or restaurants) rely on such online platforms for attracting customers. Thus, receiving less exposure than what is expected can be unfavorable for businesses. It could have a negative impact on their revenue and potentially lead to economic starvation or even shutdown.
By gathering and analyzing data from three popular platforms, we observe that many top-rated hotels and restaurants get less exposure vis-a-vis their quality, which could be detrimental for them. 
Following a meritocratic notion, we define and quantify such exposure disparity due to location-based searches on these platforms.
We attribute this exposure disparity mainly to two kinds of biases --- \textit{Popularity Bias} and \textit{Position Bias}. Our experimental evaluation on multiple datasets reveals that although the platforms are doing well in delivering distance-based results, exposure disparity exists for individual businesses and needs to be reduced for business sustainability.
\end{abstract}

\begin{IEEEkeywords}
Location-based Search, Local Recommendation, Near Me, Meritocratic Fairness, Popularity Bias, Position Bias
\end{IEEEkeywords}

%% file: intro.tex
\section{Introduction}
\label{section:introduction}
Location-based search and recommendation have become very popular nowadays, where a user can find businesses (e.g., hotels, restaurants, bars, cafes, or shops) based on their physical location. The query is often made using the user's current location or some other location explicitly entered by the user. The proliferation of smartphones with GPS availability has made these services very convenient and fast. Consequently, $30\%$ of all mobile searches today are location-based, and every month people visit around 1.5 billion destinations using location-based results\footnote{\label{note1}\url{https://www.thinkwithgoogle.com/consumer-insights/mobile-search-trends-consumers-to-stores}}. Since 2011, \emph{``near me''} or location-based searches have increased by an astounding  $3,400\%$\textsuperscript{\ref{note1}}. In the case of local recommendations, the search query is implicit and the recommendations are based on the user's inferred location.

Multiple online platforms and mobile applications like Google, Yelp, etc. allow location-based searches and recommendations in the form of \textit{nearby} feature to query restaurants, shops, hotels, etc. nearby. Moreover, hotel booking platforms like Booking[dot]com or Expedia allow travelers looking for hotels to explicitly specify a particular area of a city, and as a search result, hotels near that particular location are shown. Our analysis over a publicly available dataset by a hotel booking platform  Trivago\footnote{\url{https://recsys.trivago.cloud/challenge/dataset}} reveals that $26\%$ of users use a particular point of interest in a city to search for hotels at their destination, while the remaining $74\%$ of users use generic city names as queries. Many other online platforms like TripAdvisor, Zomato, Airbnb, etc.\footnote{\url{www.tripadvisor.com}, \url{www.zomato.com}, \url{www.airbnb.com}} also feature location-based searches and recommendations. 

With many people relying on such online platforms to choose places for their offline experiences\textsuperscript{\ref{note1}}~\cite{haoshengLocation}, these places, in order to be successful, need to get proper exposure for getting the attention of the users. Specifically, they need to appear in the top section of search results to gain attention, as users are susceptible to \textit{position bias}, which makes them pay the most attention to the top results only~\cite{craswell2008experimental,ursu2018power}. Appearing in lower ranks of search results usually translates to lower exposure, % to the users,
thereby causing loss of business opportunity for these places. Moreover, lack of proximity to popular locations, which is termed as \textit{Popularity Bias}~\cite{ciampaglia2018algorithmic, steck2011item}, can also hamper the exposure different establishments get. This is especially undesirable if the received exposure does not corroborate with the intrinsic quality of these places. 

In this work, we attempt to investigate whether the above concern actually gets manifested in real-world online platforms. By gathering and analyzing data from three popular platforms --- Google, Yelp, and Booking[dot]com, we observe that many of the top-rated establishments indeed receive lower exposure compared to their lower-rated counterparts, due to their (lack of) proximity to popular locations. 
In the offline scenario, it is well-known that location impacts the success of different businesses, and based on footfalls estimated by market research agencies, businesses often try to offer superior services and other incentives to overcome the location factor~\cite{erkucs2016innovative}. Moreover, different users have different tolerances to distance --- some users may prefer to go to a farther place to enjoy superior quality, whereas, for some other users, the distance may be a strict constraint. Present location-based search and recommendations do not consider these nuances, and ranking different places solely in the ascending order of their distances can create the unintended exposure bias. 

Our work doesn't question the idea of providing nearby establishments as results for location-based searches, rather it questions the ranking of those nearby establishments in the sole order of distance. This means that, even though the location is an important factor to determine the prosperity of an establishment, it should not be the only factor affecting its economic success. Establishments should ideally have the opportunity to compensate for a bad location with better quality, and the platforms should factor this into the ranking, thereby providing the opportunity to achieve more exposure. In future work, we plan to devise algorithms to possibly reduce the exposure bias by re-ranking search and recommendation output to effectively balance the trade-off between fairness and user satisfaction.

%% file: related.tex
\section{Related Work}\label{related}
In recent times, virtually all types of online markets have observed a colossal increase in the number of users~\cite{Danziger2019}.
While there are concerns about these platforms behaving explicitly or implicitly biased manner, they undoubtedly exert a great deal of control over how their users behave and which decisions they make.
With the possibility of the existence of bias, discrimination, and fairness concerns on these platforms, researchers have started investigating such online markets~\cite{lin2015home,levy2017designing,suhr2019two,patro2020fairrec,patro2020incremental}.
We study the literature along three dimensions:
% first, we talk about generating the top-K best recommendations following a notion of fairness; followed by a discussion on the existing biases in online platforms, and, finally, we discuss the potential fairness mechanisms in this context.
first, we talk about the existing biases on the online platforms; followed by potential mechanisms of fairness and generating the top-K best recommendations following a notion of fairness.

\subsection{Bias in the Web} 

The sheer abundance of data have lead to the compromise of its quality while extracting it from its source. A severe quality issue is data bias, which has been studied by several researchers. 
The existence of  biases affects  machine learning algorithms designed to enhance the user experience, eventually exacerbating the outcome of these algorithms, particularly in the context of recommendations and personalization systems~\cite{baeza2016data}.
There is extensive prior work that has explored various forms of biases in social media~\cite{chakraborty2016dissemination,chakraborty2017makes,Abdollahpouri2020}, polarization in the society~\cite{dandekar2013biased}, and in linguistics~\cite{otterbacher2015linguistic}. Evidence that Google's search algorithm along with its auto-complete function systematically advocates information which is either false or inclined to an extreme right wing bias, on a diverse range of subjects, has also been reported~\cite{google_bias}.
Furthermore, Nunez' article~\cite{fb_bias} indicates instances of Facebook suppressing new stories of interest to conservative readers from the social network's influential ``trending'', eventually giving rise to biases.
However, most literature  focuses on the existence of biases in social media platforms.
Our contribution lies in identifying the exposure bias existing on the sharing economy platforms.
A high correlation between online exposure of the results generated on these sharing economy platforms and their offline popularity reinforces the fact that it is important to diagnose and mitigate such biases in exposure to ensure conformity between their online received exposure and their actual quality.

\subsection{Fairness}

Data bias is inadvertently introduced into  data-driven decision-making systems, as these systems operate by learning from historical decisions, often taken by humans~\cite{speicher2018unified,zafar2017fairness}.
The  goal of such systems should be to achieve maximum utility by minimizing the errors or misclassifications over  given historical data.
However, in such data, bias is already manifested and it could be  that an optimally trained classifier does not make an impartial decision for all social groups and thereby introducing a notion of unfairness in decision making. Zafar et al.~\cite{zafar2017fairness} address this unfairness in biased data in terms of misclassification rates, as disparate mistreatment and introduces intuitive measures of this disparate mistreatment for decision boundary-based classifiers with a slight compromise on accuracy.
Another possible approach for the aforementioned problem could be by quantifying the measures of algorithmic unfairness~\cite{speicher2018unified}.
% \ashmi{TODO: check this with Speicher!}

% Fairness can also be studied under the lights of gender shades and phenotypic subgroups to evaluate bias present in automated facial analysis algorithms and datasets through Boulamwini et al.'s computer vision approach~\cite{buolamwini2018gender}.
% to bring out the significant disparities while accurately classifying different racial and gender features of the population demanding urgent attention for the creation of genuinely fair, transparent and accountable
% facial analysis algorithms.
% The notion of fairness can be extended to recommender systems both in the context of collaborative recommender systems~\cite{yao2017beyond} as well as a multi-sided concept, where  fair outcomes for multiple individuals are considered~\cite{burke2017multisided}.

%Burke is one o study of fairness in the notion of recommender systems ~\cite{burke2017multisided}
Yao and Huang illustrate a notion of fairness in the context of collaborative recommender systems~\cite{yao2017beyond}.
Experiments using their metrics, addressing different forms of unfairness, on various real data further establishes their claim of their method performing better than the baseline approaches.
On the other hand, recent works~\cite{suhr2019two,burke2017multisided,patro2020fairrec,patro2020incremental,patro2020towards,Mondal2020TwoSidedFI} bring to light the fact that fairness may be a multi-sided concept, in the context of recommendations, where the outcome should be fair for multiple stakeholders.
Depending on these considerations, several fairness-aware recommender systems as well as suggestions for possible fairness-aware recommendation architectures have been proposed~\cite{suhr2019two,burke2017multisided,patro2020fairrec,patro2020incremental,Mondal2020TwoSidedFI}.
% Speicher et al.~\cite{speicher2018unified} introduces a rather different approach to fairness by defining satisfactory measures of algorithmic unfainess. They propose a general framework using the existing inequality indices from economics to measure how unequally the outcomes of an algorithm affects different individuals or groups in a population. They provide a comparative study of (un)fairness of algorithmic predictors, thus  quantifying unfairness both at the individual
% as well as the group level.

The study by Dwork~et~al.\ introduces the notions of individual and group fairness~\cite{dwork2012fairness}.
This is analogous to our approach, where we claim that items with similar quality metrics should receive similar attention on the sharing economy platform. 
Even though Chakraborty~et~al.~\cite{chakraborty2017fair} also emphasize on a novel framework regarding fairness in the matching mechanisms of online sharing economy platforms from the providers'  perspective, the novelty of our work lies in the fact that we %recommend a re-ranked list of 
analyze the results on real-world platforms taking into consideration the inherent position bias of results and the popularity of the location query.

% A possible and relevant scenario for fairness in classification could be where individuals are classified (e.g.,admitted to the university), with the end goal to prevent discrimination against these individuals based on their affiliation to a particular group, such that the utility of the classifier (here, university) is preserved ~\cite{dwork2012fairness}. This reinforces the notion that similar individuals should receive similar treatment, which is analogous to our approach, where we claim that establishments with similar quality metrics should receive similar attention on the sharing economy platform. 

\subsection{Top-K Fair Recommendations}
The prolific need for personalized recommendations have led to the emergence of recommender systems where a filtering strategy is used to rank items that might be of interest to the user~\cite{deshpande2004item,chakraborty2019equality,Zehlike2017FAIRAF}. 
% As opposed to traditional information retrieval systems, personalized recommender systems have emerged where a filtering strategy is used to rank items that might be of interest to the current user~\cite{deshpande2004item}. 
Thus, the result of the recommendation is not a single item, but a ranked list of length K, where the user still has the opportunity to choose the final item to be consumed.
In multi-stakeholder platforms, where there are several parties involved, several notions of fairness need to be addressed by a recommendation algorithm~\cite{burke2017multisided,Mondal2020TwoSidedFI}.
%Deshpande and Karypis~\cite{deshpande2004item} proposed a particular class of model-based top-N recommendation algorithms that uses the similarities between a variety of items to identify the set of items to be recommended. 
% % The fundamental steps involved in this model include a method to compute the similarity between different items followed by another method to combine these similarities in order to compute the similarity between a basket of items and a candidate recommender item.
To achieve fairness, Chakraborty et~al.~\cite{chakraborty2019equality} envisioned top-K non-personalized crowd-sourced recommendations, such as the outcome of a multi-winner election which is periodically repeated, while selecting items at a particular time from popular websites, like Twitter, Yelp, TripAdvisor, or NYTimes~\cite{chakraborty2019equality}.
The Fair Top-K Ranking problem of selecting a subset of K items from a large pool of N candidate items by maximizing the utility (i.e., select the best" candidates)  can also be performed by subjecting to group fairness criteria such that it is capable of mitigating biases in the representation of an under-represented group along a ranked list~\cite{Zehlike2017FAIRAF}. 
% Our problem is similar to the extent that it tries to measure the user satisfaction achieved by ranking or permutation of subjects according to user preferences and then compares it with an ideal permutation of the user satisfaction.

%% file: data_gathering.tex
\section{Datasets}\label{section: data gathering}

The experimental evaluation has been carried out on three datasets obtained by querying  for businesses --- hotels and restaurants --- from popular locations on  Booking[dot]com, Yelp, and Google Places.
This situation is analogous to generating a simulation of searching for the nearby establishments from the current location of a particular user.
To demonstrate our analyses, we chose  New York City~{\textbf(NYC)} and San Francisco~{\textbf(SF)}, as these cities are both very large, and sufficiently different in terms of their neighborhood structure.
We look into three platforms --- Yelp and Google Places for the restaurant searches, and Booking[dot]com for accommodation searches.
The rankings were constructed using the order in which each result appeared in the respective searches and we use the mean ratings as a measure of the quality of each establishment. 

\subsection{Location Queries}\label{location queries}

We used publicly available Foursquare check-in information of respective 37,470 (SF) and 572,338 (NYC) total check-ins at different venues, comprising of the geographic coordinates and timestamps~\cite{yang2014modeling, yang2019revisiting}.
%Foursquare provides personalized recommendations regarding the places of interest near the user’s current location using its browsing history, nearby attractions, previous purchases, or check-in history.
We used these check-ins as proxies for user interest locations, and use the check-in counts as an approximate measure of the popularity of places in a city where most location-based searches or nearby searches are made.

We clustered the Foursquare check-ins using K-Means clustering~\cite{macqueen1967some} and with a value of K=1000, thus, having the mean cluster radii for NYC to be 499 meters and 199 meters for SF.
The cluster centroids obtained for both cities and a list of popular landmarks were used as location queries for generating results on the said platforms.
We used 561 landmarks in New York\footnote{NYC Open data: \url{https://opendata.cityofnewyork.us}}, and 272  landmarks in San Francisco \footnote{Wikipedia page: \url{https://en.wikipedia.org/wiki/List_of_San_Francisco_Designated_Landmarks}}.
The location queries dataset for both cities contained the location of each query, followed by the name of the place and its address as on Google Maps.
Here, the term location query and search query is used interchangeably within the purview of the work.

\begin{figure*}[ht]
	\centering
	\begin{subfigure}[b]{\textwidth}
		\centering
		\begin{subfigure}[b]{0.4\textwidth}
			\includegraphics[width=\textwidth]{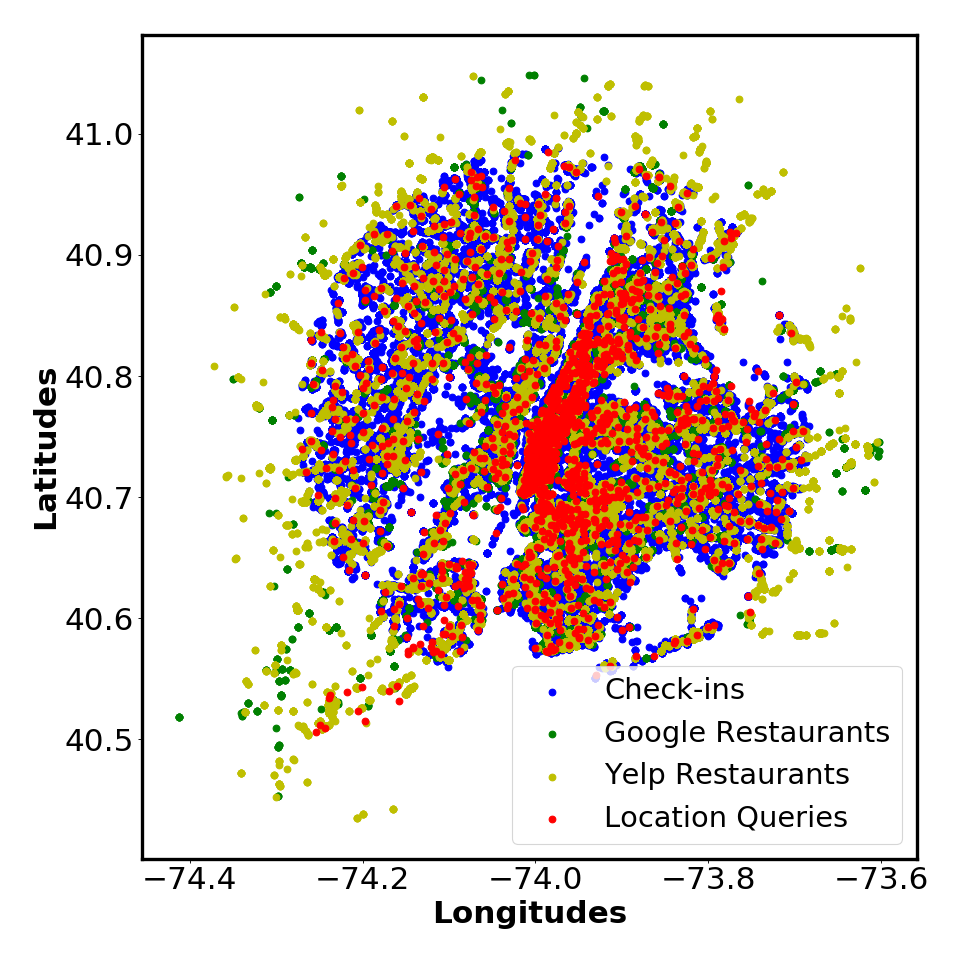}
			\caption{NYC: Google \& Yelp}
		\end{subfigure}
		\hfil
		\begin{subfigure}[b]{0.4\textwidth}
			\includegraphics[width=\textwidth]{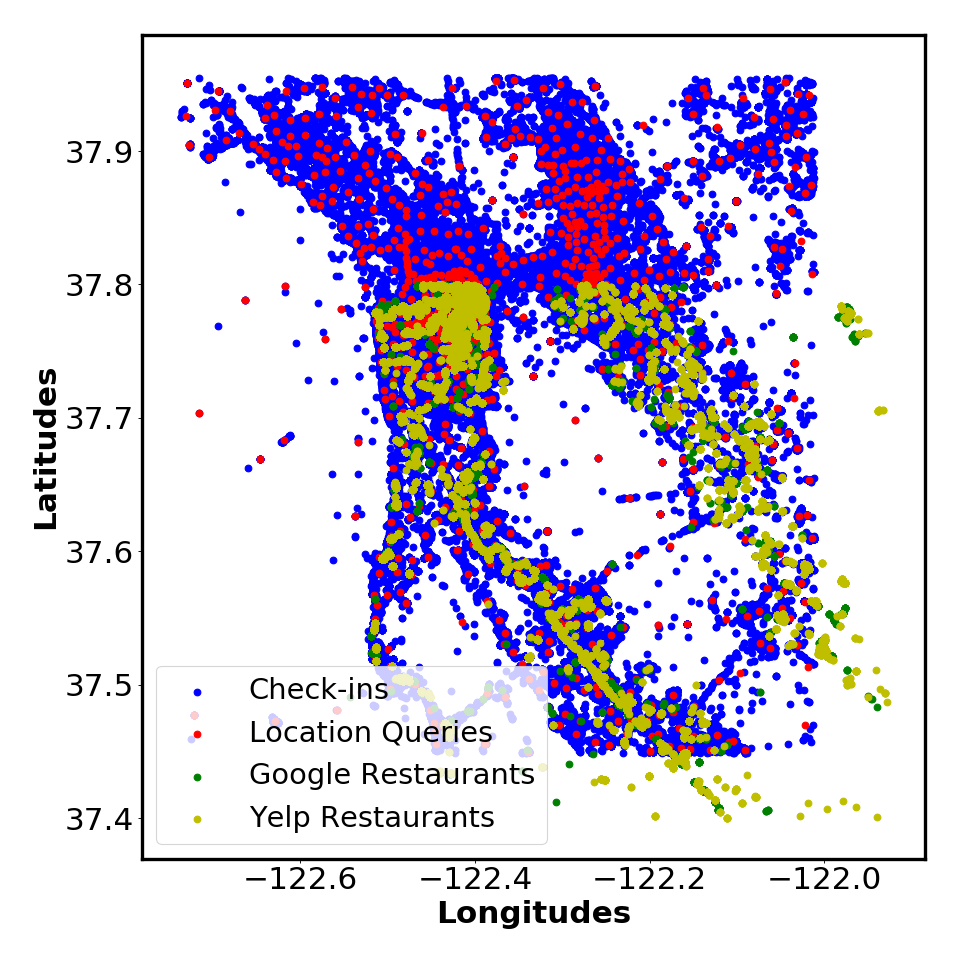}
			\caption{SF: Google \& Yelp}
		\end{subfigure}
	\end{subfigure}
	\begin{subfigure}[b]{\textwidth}
		\centering
		\begin{subfigure}[b]{0.4\textwidth}
			\includegraphics[width=\textwidth]{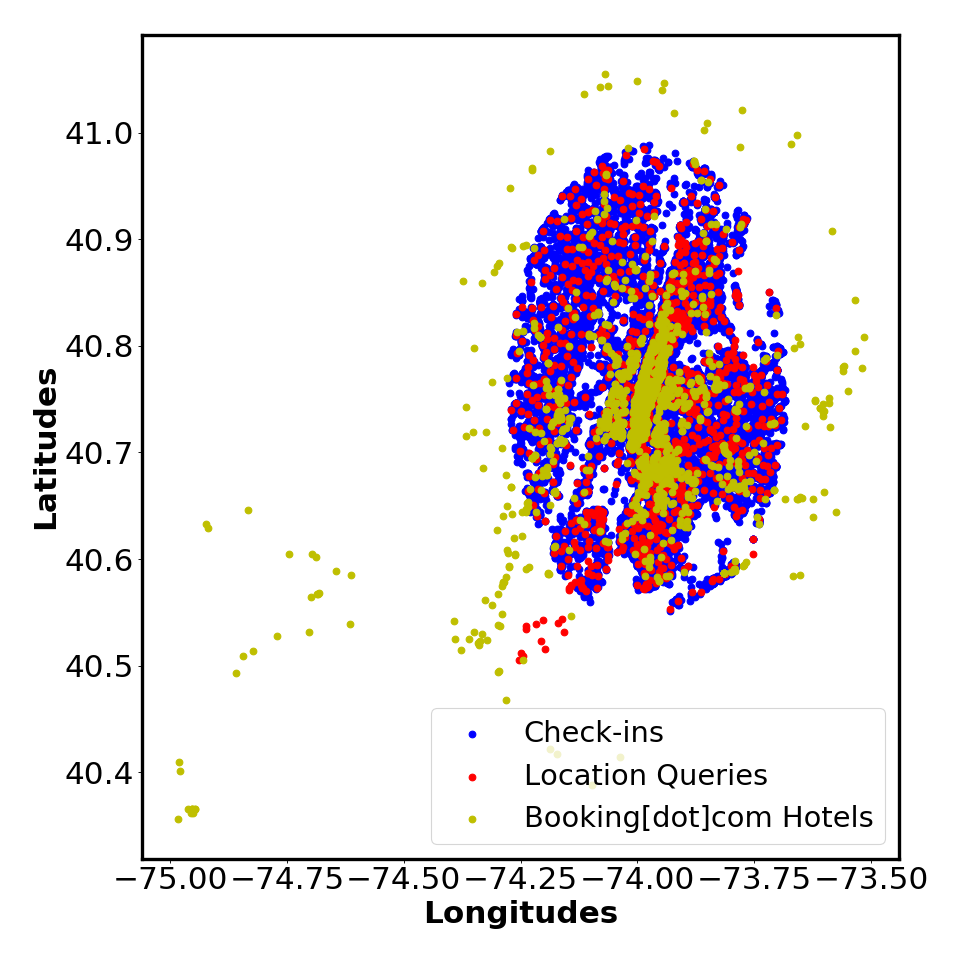}
			\caption{NYC: Booking[dot]com}
		\end{subfigure}
		\hfil
		\begin{subfigure}[b]{0.4\textwidth}
			\includegraphics[width=\textwidth]{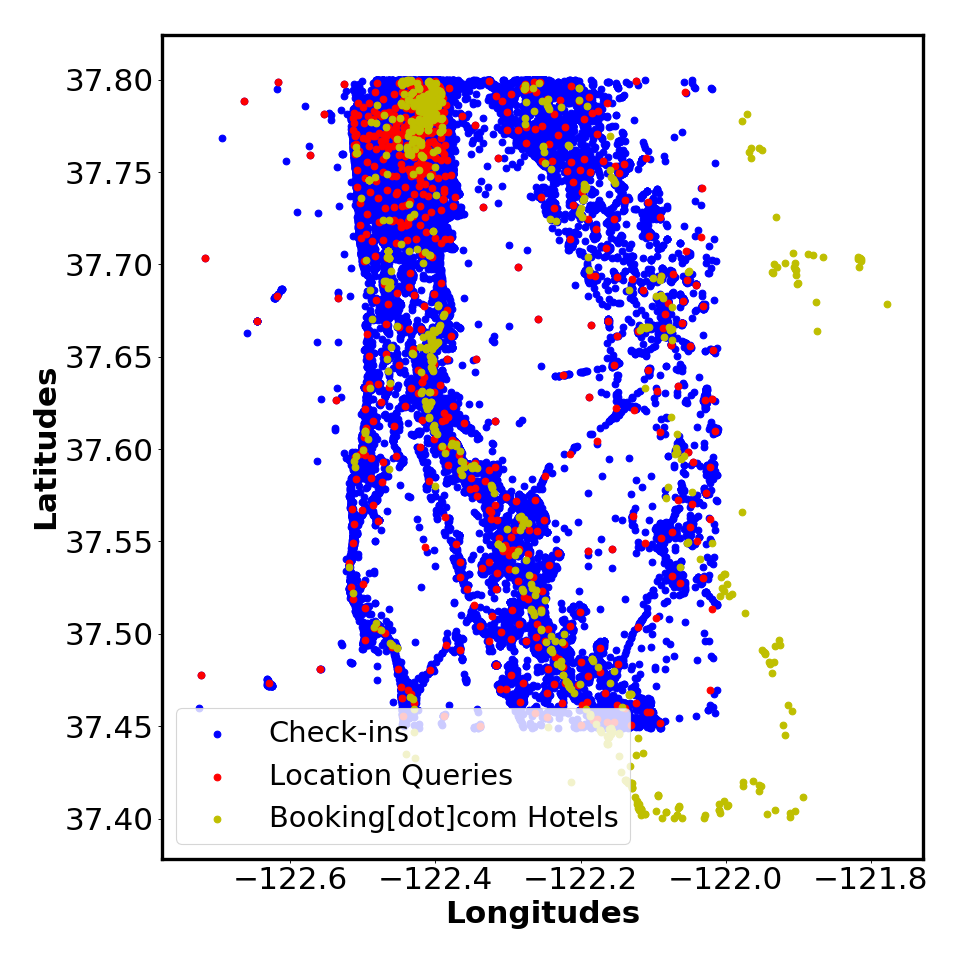}
			\caption{SF: Booking[dot]com}
		\end{subfigure}
	\end{subfigure}
	\vspace{-4mm}
	\caption{\bf Figure (a) and (b) show the distribution of check-ins, location queries, and establishments for NYC and SF respectively on Google and Yelp. Figure (c) and (d) show a similar distribution for Booking[dot]com. Each point is a coordinate represented by longitude and latitude showing the locations of check-ins, location queries, and establishments on the platforms. It can be seen that our location queries, which represent the cluster centroids, are spread across the two cities. We obtained average cluster radii of 499 meters for NYC and 199 meters for SF.}
	\label{fig:overall-distri}
	% \vspace{-4mm}
\end{figure*}

We assigned the check-ins to their nearest location query to calculate the popularity of each location query.
Haversine distance~\cite{sinnott1984virtues} was used to compute the distance between the check-ins and their nearest location query.
The popularity of a location query was calculated as the fraction of check-ins mapped to the location query to the total number of unique check-ins in the city.
We use this popularity score later as the weighing measure in our scoring model to calculate the exposure scores assigned to each establishment.
\autoref{fig:overall-distri} shows the overall distribution of the establishments with respect to the check-ins and location queries on the different platforms for NYC and SF.
It is evident from~\autoref{fig:overall-distri} that our data is distributed all over the two cities.

\subsection{Yelp}\label{subsection: yelp data}
Yelp is a platform powered by a crowd-sourced review forum which lists local places.
We used the official Yelp Fusion API\footnote{\url{https://www.yelp.com/developers/documentation/v3}} to query the different locations and collect the ranked results for restaurants.
A total of 5,821 restaurants in NYC and 9,929 restaurants in SF were obtained in this way.

\subsection{Google Places}\label{subsection: google data}
Using the aforementioned location queries, we also gathered the nearby search results on the Google Places API.
The first 20 results spanning over a radius of 8km were considered for every location query.
A total of 31,220 restaurants in NYC and 56,541 restaurants in SF could be extracted in this way.
This covers 1,761 and 4,252 unique restaurants in NYC and SF, respectively.

\subsection{Booking[dot]com}\label{subsection: booking data}
Booking[dot]com\footnote{\url{https://www.booking.com}} is a meta-search engine for making lodging reservations all across the world.
We used the results displayed on the first 10 pages of Booking[dot]com for the location queries.
Each page displayed a minimum of 15 search results. 
%Information regarding the results list displayed on the first 10 pages of the website for every search query  was parsed and combined together for the entire city.
This way, we obtained 1,547 unique lodging establishments for NYC and 4,334 for SF between 25th December 2018 and 31st December 2018.

\subsection{Dataset Statistics}
% \textbf{Basic Statistics:}
In this section, we illustrate the relevant statistics of the datasets that were gathered.
\autoref{table: basic stats} summarises the information gathered in the previous sections for a better overview of the datasets.

% \ashmi{add text}
\begin{table}[h!]
	\setlength\tabcolsep{6pt} % default value: 6pt
	\begin{tabularx}{\columnwidth}{@{} Z *{7}{c} @{}}
		\toprule 
		{} & \multicolumn{2}{c}{Yelp}  & \multicolumn{2}{c}{Google}  & \multicolumn{2}{c}{Booking[dot]com}\\
		\cmidrule(lr){2-3} \cmidrule(l){4-5} \cmidrule(l){6-7}
		& NYC & SF & NYC & SF & NYC & SF \\
		\midrule 
		\# Search Queries    & 1560 & 899 & 1560 & 899  & 1560 & 899 \\
		\cmidrule{1-7}
		\#Unique Establishments & 4554 & 3587 & 827 & 5702 & 1547 & 4334 &    \\ 
		\cmidrule{1-7}
		Minimum Rating  & 1.50 & 1.00 & 1.00 &  2.30 & 4.10 & 4.30  \\ 
		\cmidrule{1-7}
		Mean Rating     & 4.20  & 4.19 & 3.91 & 3.87 &  8.35 & 7.76 \\ 
		\cmidrule{1-7}
		Maximum Rating & 5.00 & 5.00 & 5.00 & 5.00 & 10.00 & 10.00 \\ 
		\bottomrule 
	\end{tabularx}
	\caption{\bf Basic statistics of the gathered datasets for the three different platforms for each of the two cities --- NYC and SF.}
	\label{table: basic stats}
\end{table}

\subsection{Exploratory Analyses}
%\todo{Change subsection name}
% \textbf{Ranks vs Average Distance Analysis:}
% \ashmi{rewrite text}

\begin{figure*}[ht]
	\centering
	\begin{subfigure}[b]{\textwidth}
		\centering
		\centering
		\begin{subfigure}[b]{.32\textwidth}
			\centering
			\hspace{14mm}
			\includegraphics[width=\textwidth]{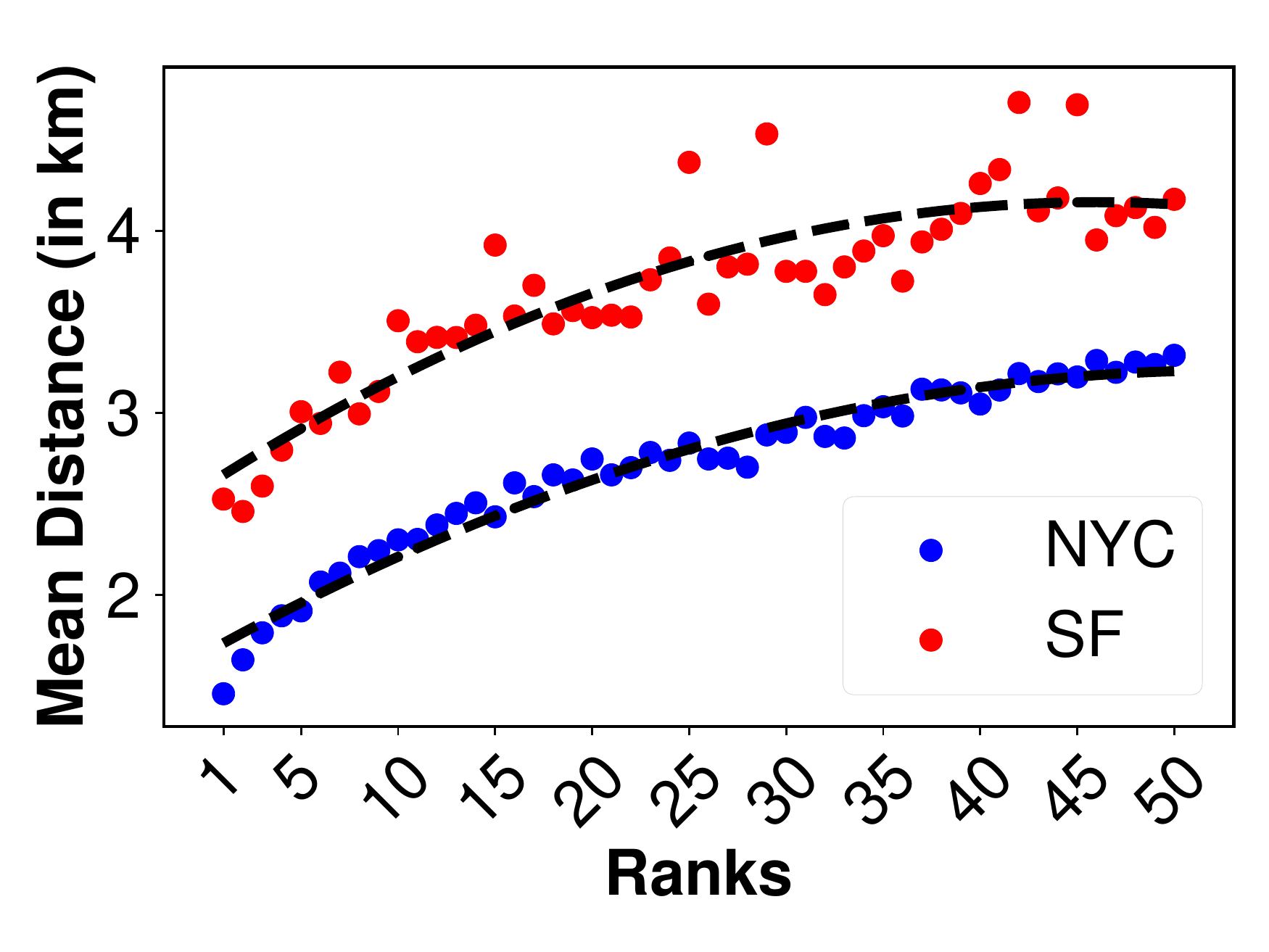}
			\caption{Yelp}
		\end{subfigure}
		\begin{subfigure}[b]{.32\textwidth}
			\centering
			\hspace{11mm}
			\includegraphics[width=\textwidth]{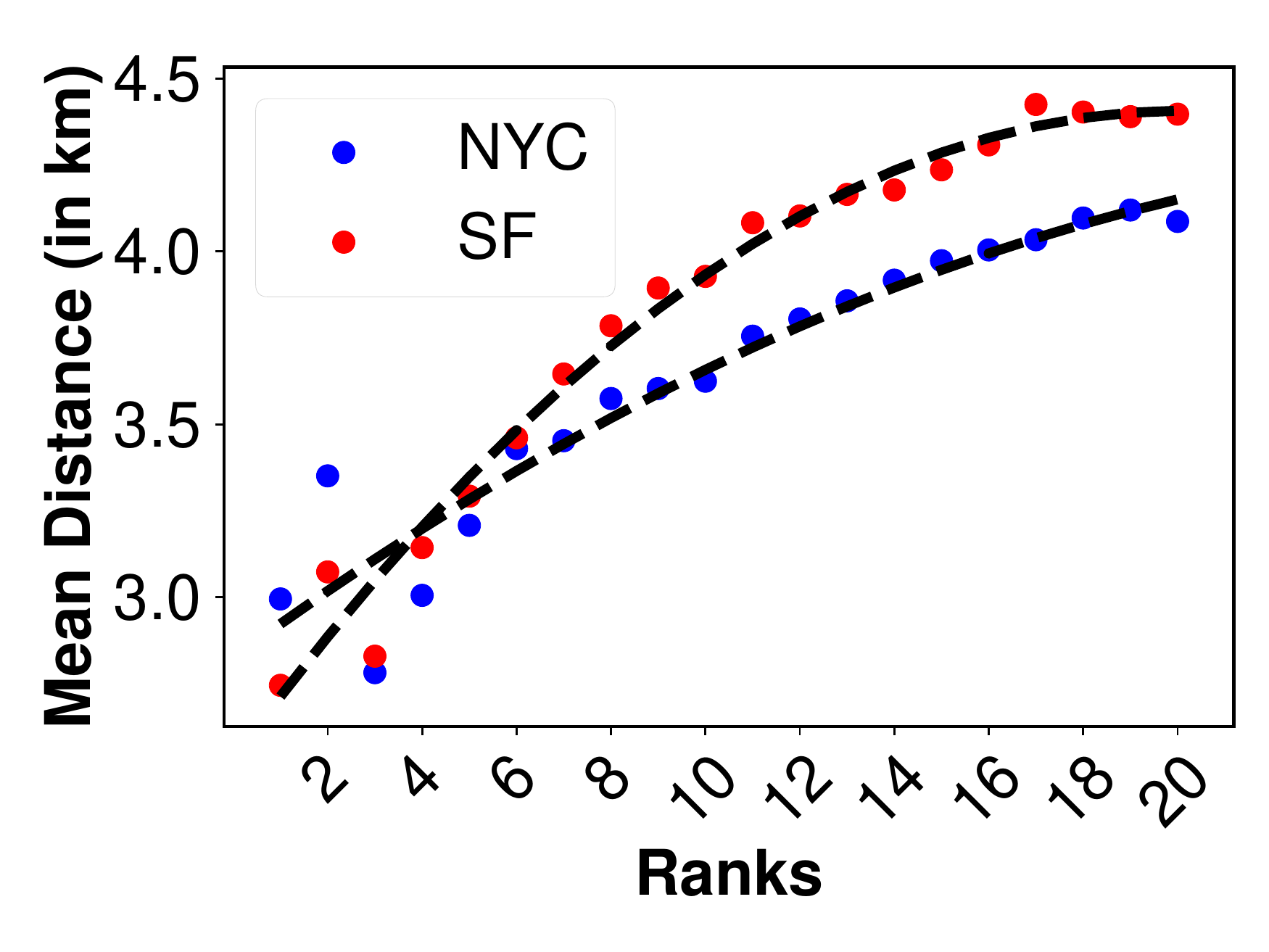}
			\caption{Google}
		\end{subfigure}
		\begin{subfigure}[b]{.32\textwidth}
			\centering
			\hspace{9mm}
			\includegraphics[width=\textwidth]{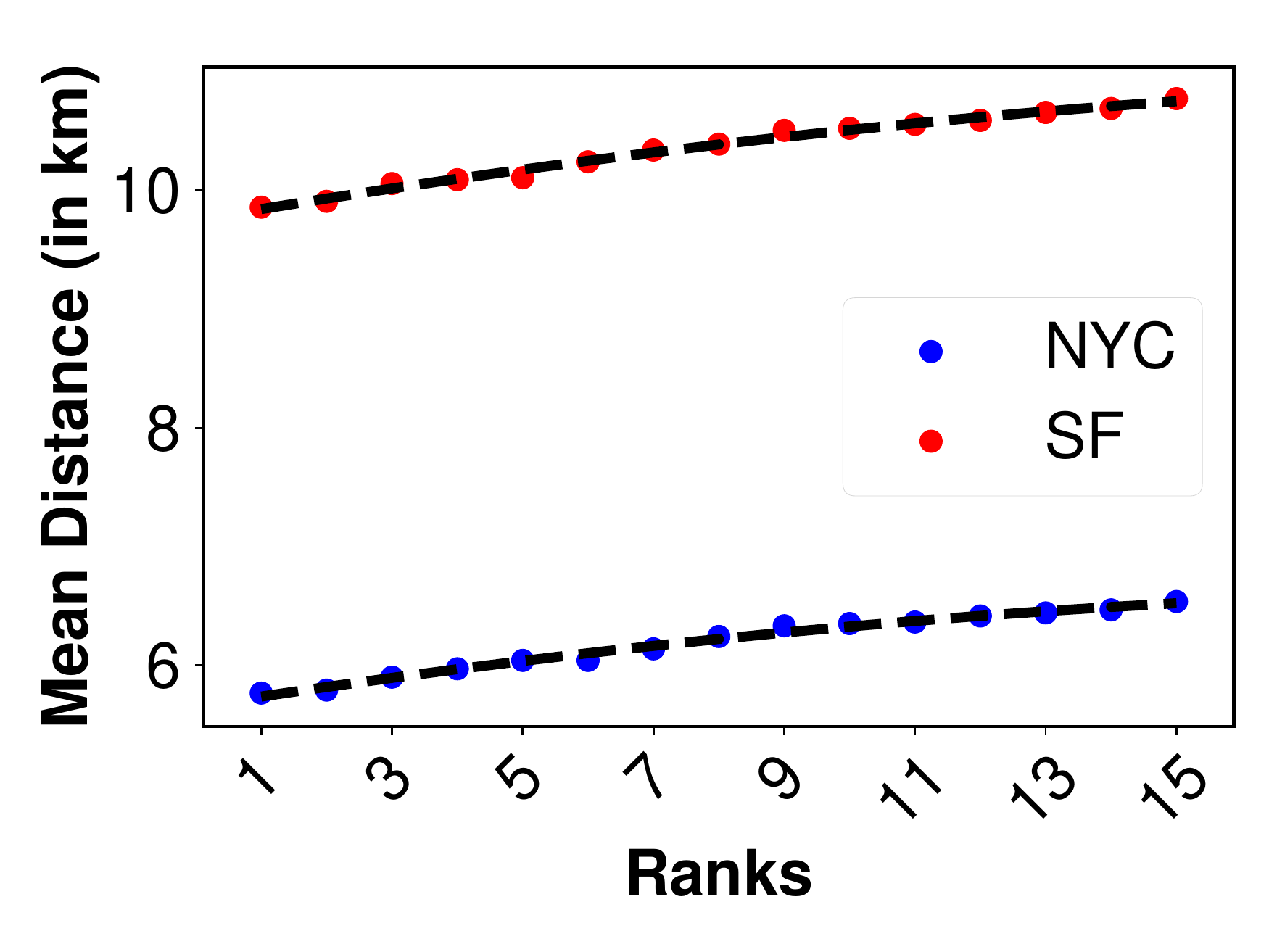}
			\caption{Booking[dot]com}
		\end{subfigure}            
	\end{subfigure}
	
	\caption{\small{\bf Ranks vs. average distance from the location queries for all three platforms. Rank here denotes the mean of the various positions at which a particular establishment appeared during all the searches. An increase in the mean distance from the location query has been observed with the increase in the ranks of the establishments. This strengthens the fact that the distance from the location query plays an important role in generating the results on these platforms.}}
	\label{fig: rank vs avg dist}
	% \vspace{-3mm}
\end{figure*}

Several exploratory analyses have been performed on the three aforementioned datasets to observe the underlying trends in the data.
\autoref{fig: rank vs avg dist} shows the mean distance of each establishment from the query location versus their ranks when displayed in the location-based search for three different platforms --- Yelp, Google, and Booking[dot]com. 
Here, rank denotes the position at which the establishment appeared in the search result.
The mean of the various positions at which a particular establishment appeared during all the searches has been denoted as the rank, for each establishment.

From~\autoref{fig: rank vs avg dist}, it is evident that the farther away a particular establishment is from the location query the lower it is ranked on average.
Although, it is not certain whether a platform ranks its results purely based on distance, the strong correlation suggests that the distance from the location query could have been taken as a very important feature for the ranking.
% The height of the error bars indicates the standard deviation with a confidence interval of 95 percent in~\autoref{fig: rank vs avg dist}.

~\autoref{table: datasetSummary ranks vs avg dist} brings out the ranks versus average distance (in kilometers) for the top three ranked establishments for the three platforms. The possible reason behind having higher distances in SF than NYC could be potentially attributed to the higher population density of NYC~\cite{cohen2015population} than SF.

\begin{table}[ht]
	\centering
	\begin{tabular}{@{\extracolsep{4pt}}rrrrrrr}
		\toprule   
		{} & \multicolumn{3}{c}{NYC}  & \multicolumn{3}{c}{SF}\\
		\cmidrule{2-4} 
		\cmidrule{5-7} 
		Ranks & Yelp & Google & Booking & Yelp & Google & Booking \\ 
		\midrule
		1   & 1.45 & 2.99 & 5.67 & 2.52 & 2.74 & 9.85 \\ 
		2   & 1.64 & 3.35 & 5.79 & 2.45 & 3.07 & 9.90 \\ 
		3   & 1.79 & 2.78 & 5.90 & 2.59 & 2.82 & 10.05 \\ 
		\bottomrule
	\end{tabular}
	\caption{\bf Comparison of ranks vs. mean distance (in km) from location queries on different platforms. The table shows that establishments with lower ranks tend to be farther away from the location queries.} 
	\label{table: datasetSummary ranks vs avg dist}
\end{table}

%% file: EB_in_search_and_reco.tex
\section{Exposure Bias in Search and Recommendation}

Subsequently, we introduce the concepts and methods we propose to capture and analyze biases in these platforms.
Even though, in this study, we focus on the local businesses like hotels and restaurants, our methods can be employed in any such sharing-economy platform that presents its recommendations in the form of ranked lists. 
%Although our analysis has been carried out on hotels and restaurants returned as a result of querying by different locations on various sharing-economy platforms, our methods can extend to any ranked list of results. 
To establish a generic terminology, we use the term \emph{``businesses''} for establishments.
Otherwise, the terms can be used interchangeably.

A location-based search query necessarily comprises of the location it is made from, but it can have various other parameters specifying e.g., types of places such as restaurants, bars, monuments, or specific types of restaurants pertaining to specific cuisine such as Italian or Asian restaurants as filtering criteria.
%The major part of any location-based search query is the base location or the location from where the search query is made.
%Other parts in the queries specifying types of places such as restaurants, bars, cafes, monuments or specific types of restaurants pertaining to a specific cuisines such as Chinese, Italian or Asian restaurants, or burger places,  are considered as filtering criteria.
These filtering criteria are usually binary, thus, independent of the ranking mechanism of the platform.
Hence, we ignore the filtering criteria in our analysis and consider only the base query locations as the search queries.

In the following, we define relevance and exposure in location-based searches.
We assume a city to have a limited number of landmarks based on which users perform nearby searches and that the landmarks have different popularities among the users.

%This assumption can be removed without affecting the implementation at any point.

\subsection{Relevance of Businesses}\label{subsection: Relevance of search results}

The relevance of a business represents how \emph{``worthy''} it is to be included as result to a user query.
%using a positive real number.
We can formalize relevance as a function from the set of locations $L$, and set of businesses $S$ to the set of non-negative real numbers. 
\begin{equation}
V: \big(L,S\big) \rightarrow\mathbb{R}_{\geq0}
\end{equation}

$V\small(l_i,s_j\small) $ is the normalized relevance of business $s_j$ when location $l_i$ is searched.

The output of the relevance function can be considered as an indicator for the \emph{desired exposure} of a business in a specific instance of location-based retrieval.
Typically platforms, however, devise their own function for relevance according to their goals and applications.
For our study, we use the mean rating of a business as the relevance function.
However, more complex functions can also be used to capture more complex requirements.

\subsection{Exposure from Search and Recommendation}\label{subsection: Exposure}

The exposure of a business is related to the amount of attention it receives from the users, which depends explicitly on the position it appears in search results, for any ranked retrieval system.
Let $\omega(t)$ be the attention a business gets when it appears in rank $t$, then the total exposure gained by a business $s_j$ over $k$ searches can be written as below.
\begin{equation}
E_j = \sum_{i=1}^{k}\omega(\rho_i(s_j))\label{eq:exposure}    
\end{equation}
Here, $\rho_i(s_j) $ denotes the rank of business $s_j$ in result (permutation) $\rho_i$.
Its value can be any integer from $1$ to $n$. The computation of $\omega(t)$ has been elaborated in~\autoref{subsection: position bias}.

\subsection{Fair Exposure for Businesses}\label{subsection: Fair Exposure}

Our notion of fair exposure in this work is analogous to the ideas of \emph{individual fairness} and \emph{distributive fairness}.
We follow \emph{individual fairness} proposed by Dwork~et~al.~\cite{dwork2012fairness}\ stating that \emph{``similar individuals should be treated similarly.''}
The similarity between individuals needs to be established based on similar values of a suitable metric specific to the task at hand.
Here, we consider businesses having similar relevance scores as similar businesses and we proclaim that two businesses are treated similarly if they receive similar exposure scores. 
% This idea is inspired by the concept of \emph{individual fairness} proposed by Dwork et. al. \cite{dwork2012fairness}. 
We consider \emph{exposure given by an online platform in total} as a readily available, yet limited resource.
How this exposure is distributed is up to the discretion of the platform's recommendation algorithm.
Potential distributive norms could be equity, equality, power, or need.

Following works on meritocracy~\cite{cappelen2010responsibility}, in this work, we use \emph{merit} as the distributive norm which requires the distribution of resources in proportion to the relevance of the businesses.
This idea is also inspired by the concept of \emph{``distributive fairness''} proposed by Rawls~\cite{rawls2009theory,rawls2001justice}.
In this regard, we use the term \emph{``meritocratic fairness''} for businesses to keep their exposures in proportion to the corresponding values of the relevance function.

An online platform's search or recommendation results $\rho_1, \rho_2, \ldots ,\rho_k$ for respective base locations $l_1', l_2', \ldots ,l_k'$  will be said to follow the notion of \emph{meritocratic fairness} if the following condition is satisfied.
\begin{equation}
\frac{E_i}{V_i} = \frac{E_j}{V_j} , \forall s_i, s_j \in S    
\end{equation}
where $V_j = \sum_{t=1}^{k}V\small(l_t',s_j\small)$ is the sum of all the normalized relevance scores of business $s_j$ over $k$ customer instances and  $\rho_1, \rho_2, ... ,\rho_k$ denotes the results of location-based searches on locations $l_1', l_2', ... ,l_k'$ correspondingly as provided by the platform.
The equality constraint defined above is a hard constraint, but we can also relax it by replacing it with soft constraints using any kind of tolerance formulation.

\subsection{Exposure Bias on Real-World Platforms} \label{subsection: EB in real-world platforms}

%\ashmi{need an intro para}
% Various online platforms and mobile applications provide location-based search and recommendation features in the form of a \emph{``Nearby''} feature to query close-by hotels or restaurants.
% Moreover, hotel booking platforms such as Booking[dot]com or Expedia give travelers the option to explicitly specify a particular area of a city as the basis for their query and prioritize hotels close to the query location.
% Many other online platforms like Google Local, Yelp, Zomato also feature local recommendations based on a customer's current location.
% Users tend to visit different establishments appearing in their search results, which in turn generate revenue for these establishments.

In location-based searches, establishments located in the proximity of the location query are usually observed to be ranked higher and establishments located farther are often ranked lower irrespective of their intrinsic quality.
Here, we consider the ratings of the establishments as the measure of their intrinsic quality. 
% Thus, the location impacts the success of different businesses~\cite{stahl1987therories}, and it has been shown that there are efforts based on footfalls estimated by market research agencies, to offer superior services and other incentives to overcome the location factor~\cite{erkucs2016innovative}.
As different parts of a city have very different popularity among customers, establishments located closer to the popular parts of the city get to be in the top results for a number of searches.
It has been often observed that even poor quality establishments located in proximity to popular parts of the city get much more exposure than many superior quality ones located slightly farther.
Such discrepancy between the \textit{actual exposure} and the \textit{deserved exposure} can be interpreted as \textit{Exposure Bias}.
We define deserved exposure as the intrinsic quality of the establishment which might be specific to different platforms or different applications, and might also depend on various factors including the ratings of place, but also the distance from the base location.
% In our work, we define \textit{deserved exposure} as a product of the total available exposure and the normalized rating.
The popularity of different parts of the city also varies over time, thereby adding another level of complexity in conceptualizing the \textit{exposure bias over time}. However, currently we do not consider the temporal dimensions for this work.
Moreover, the usual practice in the case of location-based searches is to sort the establishments, which are in the same vicinity, based on the distance and the difference in the popularity of different location queries.
This can cause high levels of exposure bias for some establishments and can potentially result in economic starvation or shutdown of the establishments which is something a recommendation platform should actively counteract.
Unfortunately,  little research has been done on quantifying this kind of exposure bias, despite the clear need to carefully analyze and mitigate it for the benefit of all stakeholders: users, businesses, and platforms.

%% file: characterizing_EB.tex
\section{Characterizing Exposure Bias}\label{methods subsection: defining EB}
% \subsection{Characterizing Exposure Bias}\label{methods subsection: defining EB}
\begin{figure*}[ht]
    % \centering
    \begin{subfigure}[b]{\textwidth}
    \centering
    \centering
            \begin{subfigure}[b]{.30\textwidth}
            \centering
           \hspace{14mm}
                \includegraphics[width=\textwidth]{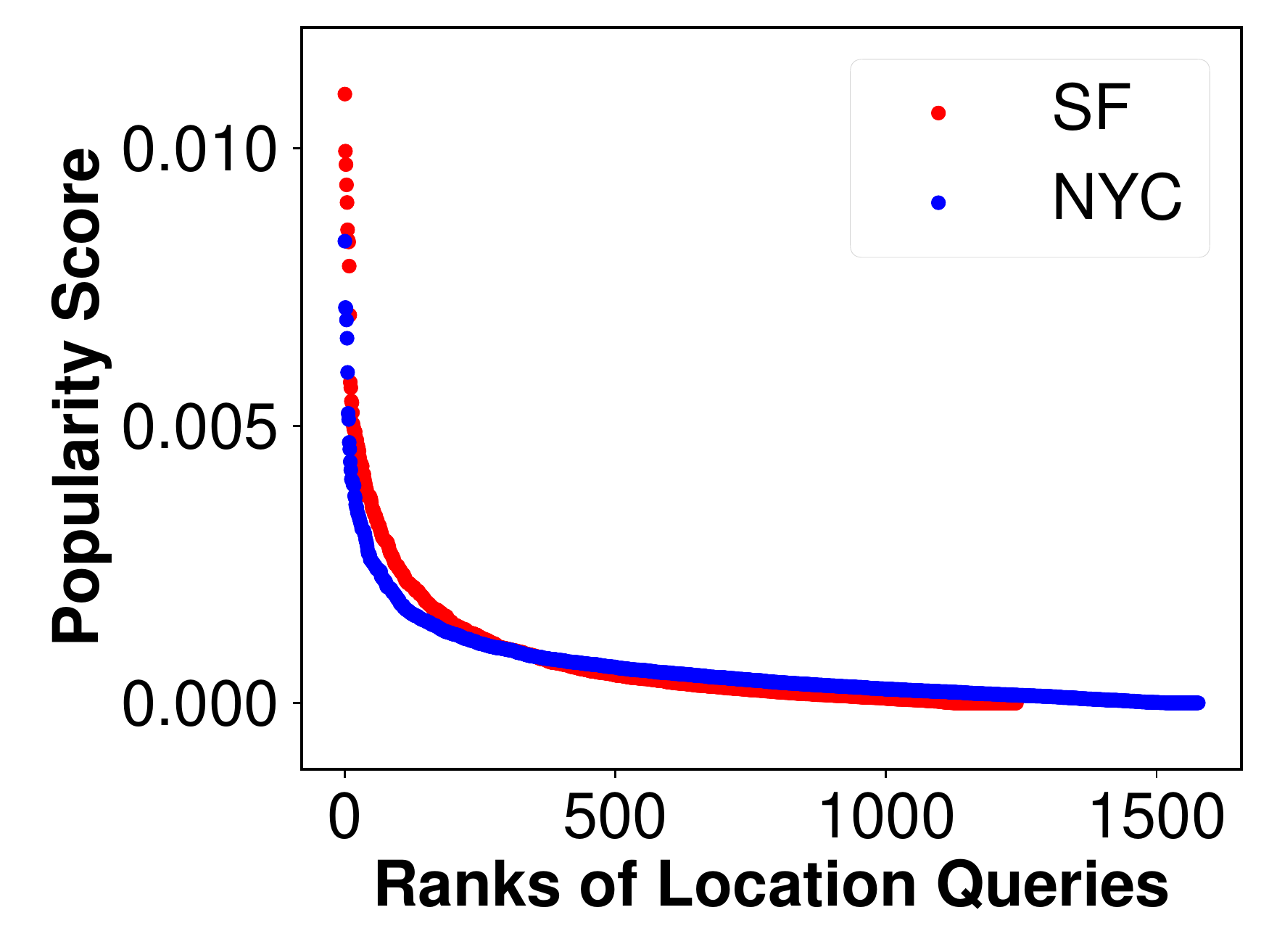}
                % \caption{\bf Popularity Distribution of different location queries across the city. It can be observed that certain areas in the city have much higher popularity than its counterparts. Therefore establishments situated closer to these popular locations tend to appear more frequently in the top results even if they do not have very high ratings. This leads to Popularity Bias where establishments located farther away from popular locations of the city are deprived of their deserved exposure.}
                \caption{Popularity Distribution}
                \label{fig: LQ popularities}
            \end{subfigure}
            \begin{subfigure}[b]{.30\textwidth}
           \centering
                \hspace{11mm}
                \includegraphics[width=\textwidth]{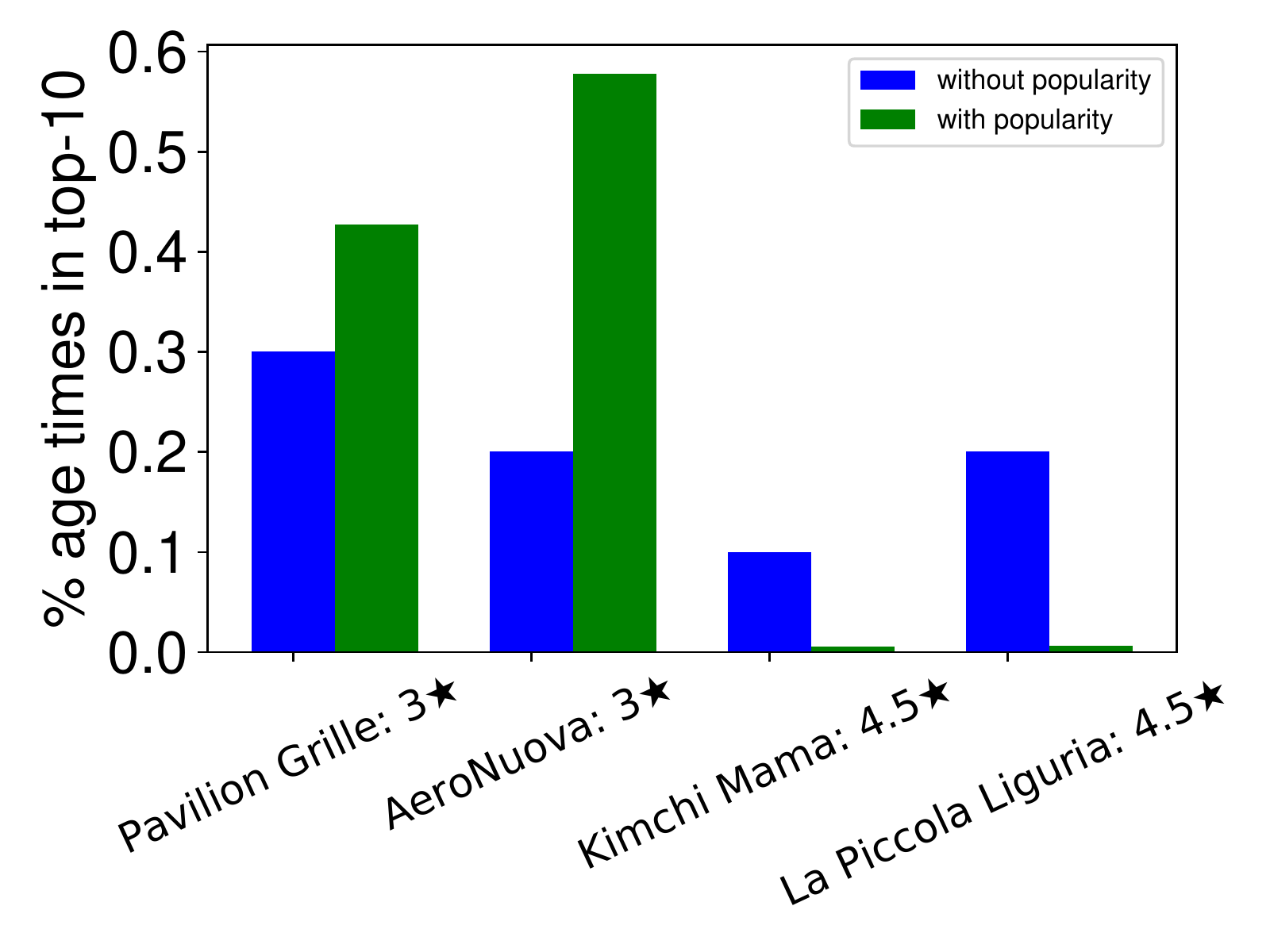}
                % \caption{\bf Examples of some top-rated restaurants from Yelp with good exposure without popularity (i.e., considering same popularity for all the location queries) versus the same when the popularity of location queries have been considered. It is evident that top-rated restaurants, little far from the popular locations end up receiving lesser exposure compared to their lower-rated counterparts.}
                \caption{Examples of restaurants}
                \label{fig: popularity vs without popularity}
            \end{subfigure}
            \begin{subfigure}[b]{.30\textwidth}
           \centering
                \hspace{9mm}
                \includegraphics[width=\textwidth]{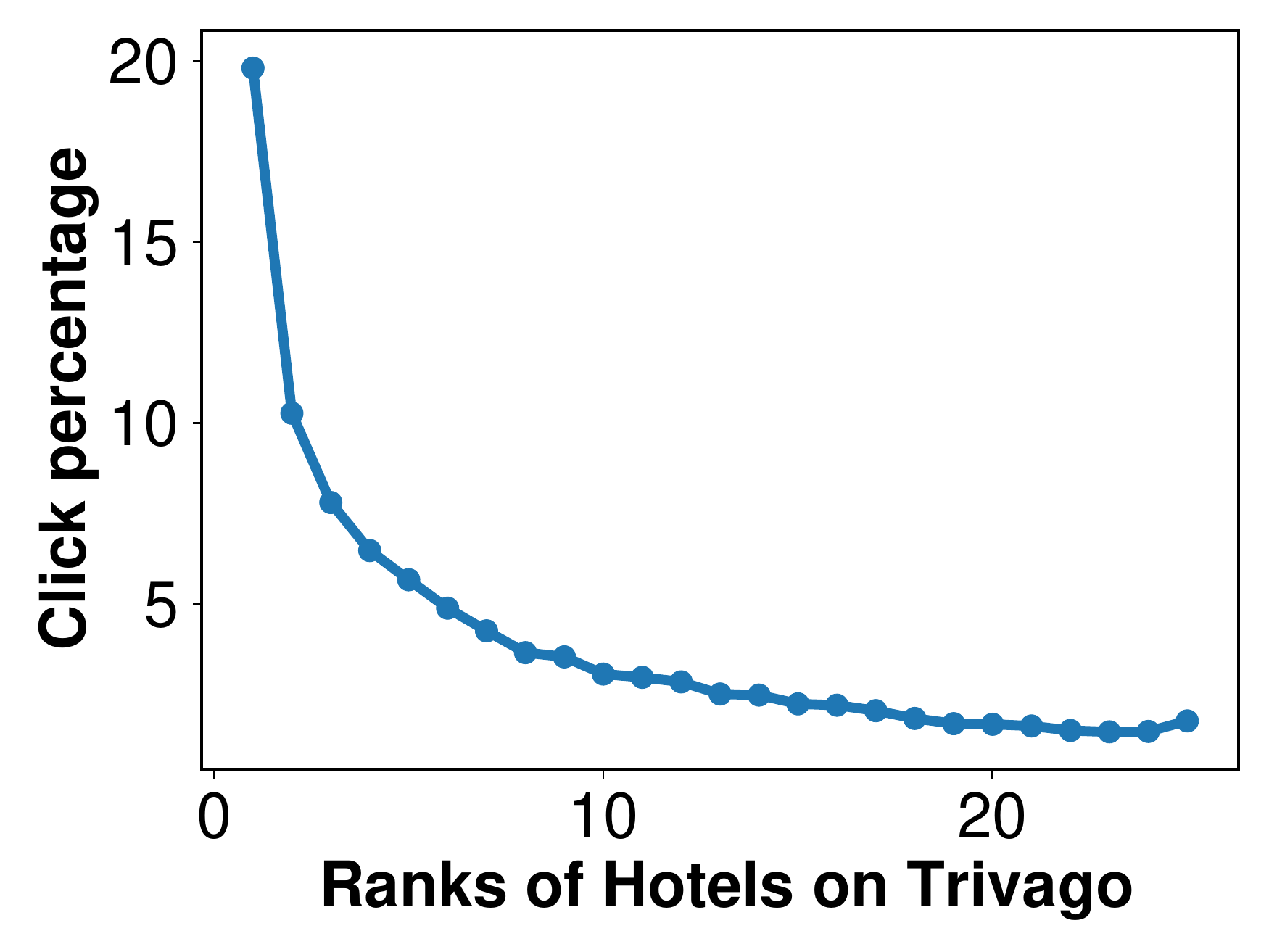}
                % \caption{\bf Trivago click-percentage (best fit p=0.144) following a geometric distribution}
               \caption{Trivago: Position Bias}
               \label{fig: trivago click-percentage}
            \end{subfigure}            
    \end{subfigure}
    % \vspace{-3mm}
    \caption{\bf Demonstration of Exposure Bias: (a) Popularity Distribution of different location queries across the city. It can be observed that certain areas in the city have much higher popularity than its counterparts. Therefore establishments situated closer to these popular locations tend to appear more frequently in the top results even if they do not have very high ratings. This leads to Popularity Bias where establishments located farther away from popular locations of the city are deprived of their deserved exposure. (b) Examples of some top-rated restaurants from Yelp with good exposure without popularity (i.e., considering the same popularity for all the location queries) versus the same when the popularity of location queries have been considered. It is evident that top-rated restaurants, farther from the popular locations end up receiving lesser exposure compared to their lower-rated counterparts. (c) Percentage of clicks on the ranked lists of results following a geometric distribution for Trivago data, bolstering the existence of Position Bias on these platforms.}
    \label{fig: characterizing EB}
    \vspace{-3mm}
\end{figure*}
Essentially, the exposure bias for any business can be described as the difference between its relevance to the query and the user attention it receives.
Hence, the exposure bias $B_j$ for business $s_j$ when query with base location $l'_i$ gives ranking $\rho_i$, can be written as below.
\begin{equation}
B_j\small(\rho_i,l'_i\small)=V\small(l'_i,s_j\small)-\omega(\rho_i(s_j))
\label{eq:exp_bias_single}
\end{equation}

The overall exposure bias can be computed in the following ways:
\begin{center}
	\begin{tabular}{cl}
		Utilitarian Measure & $^1B= \sum\limits_{j=1}^{n}|B_j|$\\
		Egalitarian Measure & $^2B= \max\limits_{j\epsilon\small[1,n\small]}|B_j|$\\
	\end{tabular}
\end{center}

In this paper, we focus on the utilitarian measure.
As it is difficult to achieve zero exposure bias in each search ranking, it is more useful to quantify the bias in cumulative form over a number of searches.
We utilize the idea of \emph{amortized fairness} as proposed by Biega~et~al.~\cite{biega2018equity} to define cumulative exposure bias here.
The exposure bias introduced by an online platform's $k$ search results $\rho_1, \rho_2, ... ,\rho_k$ for respective user locations $l_1', l_2', ... ,l_k'$, can be seen as below.
\begin{equation}
B = \sum_{j=1}^{n}|V_j-E_j| = \sum_{j=1}^{n}\Big|\sum_{i=1}^{k}\{V\small(l_i',s_j\small)-\omega(\rho_i(s_j))\}\Big|\label{eq:exp_bias_all}
\end{equation}
Results of location-based search on a platform are said to follow \emph{meritocratic fairness} if exposure bias approaches zero.

We consider position bias and popularity bias to be responsible for above exposure bias. 
We define them in the following sections, as two components of this exposure bias.
Even though position bias has been studied earlier~\cite{biega2018equity, dwork2012fairness, singh2018fairness}, fewer focus has been given to popularity bias~\cite{abdollahpouri2019managing, nematzadeh2017algorithmic}.
% The cause of popularity bias can be attributed due to the differences between the popularity of queries and the relevance-based ranking system adds up more and more exposure for items that are closely related to the popular queries.
\subsection{Popularity Bias} \label{subsection: popularity bias}

% The cause of popularity bias can be attributed due to the differences between the popularity of queries and the relevance-based ranking system adds up more and more exposure for items that are closely related to the popular queries.
By gathering and analyzing data from three popular platforms, namely Google, Yelp, and Booking[dot]com, we observe that many of the top-rated places indeed receive lower exposure compared to their lower-rated counterparts due to their lack of proximity to popular locations.
Such popularity bias has been previously studied under the context of recommender systems and social media~\cite{nematzadeh2017algorithmic, steck2011item}.
In our analyses, we seek to provide evidence for the existence of unintended exposure bias due to the popularity of location queries.

The frequencies of location queries are different in different parts of the city, with more popular and known districts and landmarks being queried more often (Figure~\ref{fig: LQ popularities}).
Thus, establishments situated near the popular locations of the city, tend to be in the top results more often.
It has been observed many times that even lower-rated establishments or establishments of poor quality, when located in the proximity of popular regions, receive much more exposure than their higher-rated counterparts, which are located slightly farther.
Figure~\ref{fig: popularity vs without popularity} shows examples of how the popularity of location queries has caused top-rated restaurants---farther from the popular locations---to end up receiving lesser exposure compared to their lower-rated equivalents. 
% lose too much of exposure when popularity of location queries is considered, and examples of some lower-rated counterparts with too much gain in exposure after considering the popularity of its location queries.
% To calculate the popularity of a location query, the Foursquare check-ins were mapped to their nearest location query using Haversine distance~\cite{sinnott1984virtues}.
% The mean exposure scores were plotted against the popularity scores of the cluster centroids, for both the cities on all three platforms.
% \ashmi{may be add a fig here?}The trend, that we observed, suggests that the exposure received by the establishments for a particular location query is directly proportional to its popularity.
% In other words, the higher the popularity of a location query, the more exposure was observed to be received by the establishments of that region.
% This could possibly attribute to the existence of popularity bias for these search queries on these platforms.

\subsection{Position Bias} \label{subsection: position bias}

Position Bias or the bias induced due to the ranking of the businesses have been studied in earlier works~\cite{biega2018equity,singh2018fairness}.
Owing to position bias, disproportionately less attention is being paid to businesses that are ranked lower in the search results obtained in a ranked retrieval system.
Our work is inspired by the work by Biega~et~al.~\cite{biega2018equity}, which focuses on the relationship between relevance and attention.
Since relevance can be imagined as a proxy for worthiness in the context of a given search task, the attention received by a business from searchers should ideally be proportional to their relevance.
Studies have shown that establishments need to appear in the top results of a search query in order to gain attention, as users are susceptible to position bias, which makes them pay the most attention to the top results~\cite{craswell2008experimental, ursuRanking}.
Appearing in lower ranks of search results usually translates to lower exposure, thereby causing loss of business opportunity for these places.
It is especially undesirable if the received exposure does not corroborate with the intrinsic quality of these places, which we approximate using their user ratings. 

Our proposed models require a weight assigned to each position of the ranked list, denoting the fraction of the total attention the position gets. The higher the position of the establishment is on the list, the higher weight will be assigned.
To calculate exposure scores, we require the fraction of attention each position in the search result gets, to be the normalized amount of users' attention to the position. These values can be calculated from the previous user behavior.

In this regard, we used a publicly accessible dataset provided by Trivago (made available during RecSys 2019 Challenge\footnote{\url{https://recsys.trivago.cloud}}) to study the user behavior on the Trivago hotel search platform.
It was observed from Figure~\ref{fig: trivago click-percentage} that the percentage of clicks on the ranked lists of results geometrically decreased with the increase in the position at which they appeared as a result of the search.
Therefore, for our analysis, we have considered the following Geometric-$\kappa$ attention distribution for assigning the exposure scores to the establishments.\\

% Following the definitions of Biega~et~al.~\cite{biega2018equity}, here we consider the following attention distributions.
\noindent {\bf Geometric-$\kappa$ Attention:}
The weights are distributed geometrically with parameter $p$ starting from position $1$ to $\kappa$. It remains 0 for positions beyond $\kappa$. The distribution of attention is given by:
	\begin{center}
		$\omega_t$ = $\Bigg\{$\begin{tabular}{cc}
			$\frac{p\small(1-p\small)^{t-1}}{\sum_{j=1}^{\kappa}p\small(1-p\small)^{j-1}}$ & $1\leq t\leq\kappa$  \\
			$0$ & $t>\kappa$  
		\end{tabular}
	\end{center}

The value of $\kappa$ has been considered to be the number of results returned on the first page of the search query on each platform. Therefore, we have considered a value of $\kappa$ equals 50, 20 and 15 for Yelp, Google, and Booking[dot]com respectively.
The value $p$ equals 0.144 has been computed from Figure~\ref{fig: trivago click-percentage} as the best fitting value for the curve.
By adding up the individual attention scores received over different user instances, we compute the total exposure for every business on a platform (as given in \autoref{eq:exposure}), and then, the total exposure bias for each business and for the particular platform (using \autoref{eq:exp_bias_single} and \autoref{eq:exp_bias_all}). 
\begin{figure*}[t!]
    % \centering
    \begin{subfigure}[b]{\textwidth}
    \centering
    \centering
        
         \begin{subfigure}[b]{.30\textwidth}
            \centering
          \hspace{14mm}
                \includegraphics[width=\textwidth]{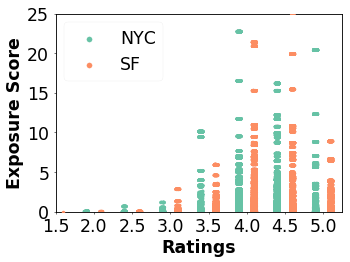}
                \caption{Yelp}
            \end{subfigure}
            \begin{subfigure}[b]{.30\textwidth}
          \centering
                \hspace{11mm}
                \includegraphics[width=\textwidth]{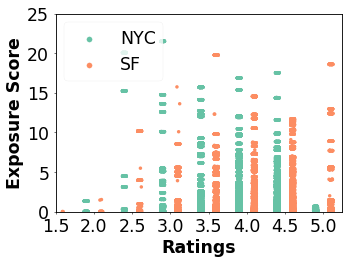}
                \caption{Google}
            \end{subfigure}
            \begin{subfigure}[b]{.30\textwidth}
          \centering
                \hspace{9mm}
                \includegraphics[width=\textwidth]{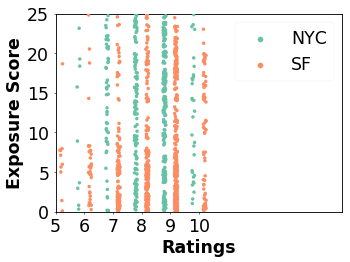}
                \caption{Booking[dot]com}
            \end{subfigure}            
    \end{subfigure}
    \vspace{-3mm}
    \caption{\bf Ratings vs. exposure scores for all three platforms --- Yelp, Google, and Booking[dot]com. It can be seen that there are several establishments on these platforms which have higher ratings but receive lower exposure and vice-versa. This further reinforces the existence of exposure bias on these platforms.}
    \label{fig: ratings vs exposure scores}
    % \vspace{-5mm}
\end{figure*}
% Exposure Bias in Location-based Search -- plots (WWW ppt)-- EB exists-- lower rated establishments have higher exposure and vice versa, show for Yelp, Booking and Google (Exposure Score Vs Ratings)
\subsection{Exposure Scores vs. Ratings}
Each of the establishments obtained from the various platforms --- Yelp, Google, and Booking[dot]com were scored against the Geometric-$\kappa$ attention mechanism scheme elaborated in section~\ref{subsection: position bias}, based on their appearance in the search result. The mean exposure score for every establishment was then computed under this scheme.  

To assess the impact of the popularity of the search query on the exposure scores, first, we assigned the scores to the establishments depending on their appearance in the search result. Then they were weighed against their popularity score. The mean of these weighted exposure scores was calculated for every establishment.
\autoref{fig: ratings vs exposure scores} shows the variation of exposure scores with respect to the establishment ratings for the Geometric Attention distribution scheme for the different platforms.
It is evident from~\autoref{fig: ratings vs exposure scores} that certain establishments despite having a high rating (4.0 or above for Yelp and Google and greater than 6.0 for Booking[dot]com) receive lower exposure than their lower-rated counterparts and vice-versa, thereby proving the existence of exposure bias on these platforms.
However, the range of exposure values across different ratings are similar indicating that it doesn't follow a meritocratic notion.

\subsection{Observed Exposure Bias}\label{results: measuring exposure bias}

Exposure bias seems to exist when low rated establishments have higher exposure than their higher-rated counterparts. In other words, when the percentage of establishments with high ratings receiving low exposure exceeds the ones with high ratings and high exposure and vice versa.
To calculate the existing exposure disparity with respect to the ratings we define the following terms ---

\begin{itemize}
  \item \textit{\bf High Exposure (HE)}: Exposure scores belonging to the top 25 percentile.
  \item \textit{\bf Low Exposure (LE)}: Exposure scores belonging to the bottom 25 percentile.
  \item \textit{\bf High Rating}: For Google and Yelp, since the rating distribution is discrete, any restaurant having a mean rating of 4.0 or higher has been considered to fall in the range of highly-rated restaurants. However, for Booking[dot]com, since the rating distribution was continuous, therefore the establishments with ratings in the top 25 percentile have been considered to be highly-rated.
  \item \textit{\bf Low Rating}: Similarly, for Google and Yelp, any restaurant having a mean rating lower than 4.0 has been considered to fall in the range of low rated restaurants. For Booking[dot]com, the establishments with ratings in the bottom 25 percentile have been considered to be low rated.

\end{itemize}

 We divided the set of establishments on the three platforms into two groups --- establishments with \textit{low ratings} and \textit{high ratings}. For each of these groups, we calculated the percentage of establishments receiving \textit{high exposure} \textbf{(HE)} and \textit{low exposure} \textbf{(LE)}. 
The whole experiment was done under two scenarios --- 
\begin{itemize}
    \item when all location queries were considered to have equal popularities (Position Bias)
    \item when the popularity of the location queries was considered (Popularity Bias)
\end{itemize}

The results of the fraction of establishments exhibiting exposure bias on the different platforms were summarized in~\autoref{table: position bias percentage comparison}. The range of exposure values across
different ratings --- both high and low are similar, indicating that it does not follow a meritocratic notion.
% The results were documented in~\autoref{table: position bias percentage comparison}. 
% \subsubsection{Measuring Position Bias}\label{results: measuring position bias}

% Position Bias is the scenario of exposure bias where the popularity of the location query remains the same for all location queries. This type of exposure bias can be computed by calculating the difference between the exposure received by the establishment and its actual relevance. This is similar to the calculation of exposure bias in section~\ref{methods subsection: EB}. The only difference is that since the popularities of all location queries are assumed to be equal, hence the exposure scores are not weighted against the popularity of the location queries.

% The table summarises the percentage of establishments showing exposure disparity owing to Position Bias. It has to be noted that for Booking[dot]com since the lower quartile and upper quartile of mean ratings determine the bound for low and high ratings respectively, there exists some establishments which lie within the inter-quartile range. However, those have been omitted here for symmetry and simplicity. 

\begin{table}[t!]
\setlength\tabcolsep{3.5pt} % default value: 6pt
\begin{tabularx}{\columnwidth}{@{} Z *{5}{c} @{}}
\toprule 
{} & {}  & \multicolumn{2}{c}{NYC}  & \multicolumn{2}{c}{SF}\\
\cmidrule(lr){3-4} \cmidrule(l){5-6} 
& Rating & LE & HE & LE & HE \\
\midrule 
Yelp    & Low  & 29.05 & \textcolor{red}{70.94} & 35.42  & \textcolor{red}{64.57}  \\ 
        & High & \textcolor{red}{10.65} & 89.34 & \textcolor{red}{11.81} & 88.18    \\ 

\cmidrule{1-6}
Google  & Low & 60.38 & \textcolor{red}{39.61} &  43.28 & \textcolor{red}{56.71}  \\ 
        & High & \textcolor{red}{51.69} & 48.30 & \textcolor{red}{39.03} &  60.96  \\ 
\cmidrule{1-6}
Booking[dot]com* & Low & 24.54  & \textcolor{red}{25.58} & 34.63 & \textcolor{red}{18.64}  \\ 
         & High & \textcolor{red}{16.00} & 32.00 & \textcolor{red}{25.00}  & 28.12  \\ 
\bottomrule 
\end{tabularx}
% \begin{tablenotes}
%       \small
%       \item *Since the lower quartile and upper quartile of mean ratings determine the bound for low and high ratings respectively, there exists some establishments which lie within the inter-quartile range. However, those have been omitted for symmetry and simplicity.
%     \end{tablenotes}
\caption{\bf Percentage of establishments showing exposure disparity owing to Position Bias. HE denotes high exposure (top 25 percentile), LE denotes low exposure (bottom 25 percentile). Low and High denote low and high ratings respectively. The values in red indicate the observed exposure disparity. The exposure disparity values indicate that exposure bias does not follow a meritocratic notion.}
\label{table: position bias percentage comparison}
\end{table}

%% file: conclusion.tex
\section{Conclusion}
We theoretically formulate the idea of exposure bias due to location-based retrieval. In our experimental study using real-world data, we show that substantial discrepancy exists between the actual exposure received by the establishments and their deserved exposure, despite several establishments having equal relevance scores. We attribute this exposure disparity to mainly two kinds of biases --- \textit{Popularity Bias} and \textit{Position Bias}. While Position Bias has been previously addressed by several studies, investigating Popularity Bias in the domain of multi-sided market platforms remains a relatively less explored problem. Our work makes the novel contribution of identifying the existence of unintended exposure bias due to \textit{Popularity Bias} and \textit{Position Bias} in location-based retrievals. To the best of our knowledge, we are the first to quantify such bias in the case of location-based retrieval.

Exposure preservation is necessary, so that establishments are not deprived from their deserved exposure both online and offline.
There is evidence that online exposure can lead to offline economic opportunity for the establishments as many people rely on online platforms for their offline experiences~\cite{haoshengLocation}. Hence, this bolsters our notion of fairness, according to which all the establishments should receive online exposure in proportion to their quality or merit. 

Our work is focused on the real-world data gathered from Yelp, Google and Booking[dot]com for two cities, namely New York City and San Francisco. However, the analysis can be extended to other cities in the world and on other  platforms such as Airbnb or Trivago as well.
Various demographic and economic influences of the neighborhoods in which these establishments are located and their impact on their exposure can also be explored in future work. 

Finally, while we have discussed the potential causes of exposure bias and established the existence of considerable discrepancy in exposure among various establishments, methods to mitigate this still remain unexplored. In future we plan to design an online optimization framework to minimize exposure bias over time and mitigate the effects of popularity bias and position bias. 
% Our experimental evaluation reveals that these platforms are  doing substantially well in exposure preservation on an overall level. However, on an individual level exposure bias still exists and is evident from the examples in~\autoref{results: measuring position bias} and~\autoref{results: measuring popualarity bias}. Thus, mitigation of exposure bias could be a possible future work and extension of this thesis.